\documentstyle[sprocl]{article}

\def\Journal#1#2#3#4{{#1} {\bf #2}, #3 (#4)}

\def\AJ{\em AJ}
\def\APJ{\em ApJ}

\def\be{\begin{equation}}
\def\ee{\end{equation}}
\def\bea{\begin{eqnarray}}
\def\eea{\end{eqnarray}}

\begin{document}

\title{STAR COUNTS FROM HST: IMPLICATIONS FOR DARK MATTER}

\author{ANDREW GOULD\footnote{Alfred P.\ Sloan Foundation Fellow}}

\address{Dept of Astronomy, 174 W 18th Ave, Columbus, OH 43210, USA}


\maketitle\abstracts{
Star counts made with the Hubble Space Telescope (HST) probe four populations
that are important for dark matter: disk, halo, bulge, and intergalactic.
The disk mass function falls for masses $M<0.6\,M_\odot$ in sharp contrast
to the rising Salpeter function usually assumed.  The amount of ``observed'' 
disk material is therefore lower than commonly believed which implies the need
for disk dark matter.  Halo stars contribute no more than a few percent of
the dark matter. Disk and halo together contribute no more than 10\% of the
observed microlensing optical depth toward the Large Magellanic Cloud.
The bulge luminosity function is similar to that of the disk down to
$M_V\sim 10$.  If this similarity continues to the bottom of the main sequence,
the bulge microlensing events can only be explained by a large population
of brown dwarfs.  Intergalactic stars in the Local Group have a density
lower than the local halo density by at least $10^{-3.5}$.
}
  
\section{Introduction}

	Counting stars is a powerful method for probing galactic structure,
but until recently it has been limited to stars $I<19$.  At fainter magnitudes
galaxies vastly outnumbers stars.  Although galaxies are typically resolved 
even in ground-based images and therefore can usually be distinguished from 
the point-like stars, some galaxies with steep surface-brightness profiles
avoid detection and pollute the sample.  The problem grows worse rapidly at 
lower flux levels since the galaxies become smaller, fainter, and more 
numerous.  Heretofore, intrinsically faint stars could therefore be studied
only when they were found nearby.  For the faintest stars, the volume probed 
was so small that measurements of the luminosity function (LF) were both
highly uncertain and highly controversial.  One result of this is that most
people have assumed that the mass function (MF), which is derived from the LF
using a mass-luminosity relation, continued with its Salpeter slope
\be 
{d N\over d\log M}\propto M^\alpha\qquad (\alpha=-1.35,\rm Salpeter)
\label{eq:salp}
\ee
as measured for relatively massive stars.  This then led to the assumption
that there was a large quantity of stellar matter which was not observed
but must ``certainly'' be there if only our instruments were powerful enough
to see them.  Hence, people would routinely quote high mass-to-light
ratios $(M/L\sim 10)$ for the luminous components of galaxies believing that
``dark matter'' was needed only to account for the rest.  In the case of the 
Milky Way disk, at least, we now have the powerful instrument at our disposal,
but we do not see the stars.  In the case of the bulge, we are able to see
much fainter than before, although we still do not probe directly the region
of the MF corresponding to the place where the disk MF turns over.  
Nevertheless, we must begin to suspect that the disk and bulge MFs are similar
and that the large mass which is dynamically determined to be associated with 
the luminous components of galaxies is not in the form of low-luminosity
stars.

\section{The Disk Mass and Luminosity Functions}

	Gould, Bahcall, \& Flynn~\cite{gbfI} identified 192 M dwarf stars in 
22 fields
imaged by the Wide Field Camera (WFC2) on HST to an average limiting
magnitude of $I=23.7$, about 100 times fainter than the limit of typical of 
ground-based surveys.  We combined these with a brighter sample of 65 M dwarfs
identified in 162 fields imaged with the pre-repair Planetary Camera.  We 
found that the LF clearly peaks at about $M_V\sim 12$ ($M_I\sim 9$).  The
transformation from an LF to an MF requires some care because the 
mass-luminosity relation is non-linear.  However, using the empirically 
measured relation of Henry \& McCarthy~\cite{hm}, we found that the MF peaks
at about $M\sim 0.6\,M_\odot$.  The detailed structure of the faint end of 
the LF remained poorly determined because there were only a total of 23 stars
with $M_V>13.5$.  However, we have now analyzed an additional 31 WFC2 fields
which contain a total of 24 stars in this faint region.~\cite{gbfII}  
We now find a clear
break in the MF at $M\sim 0.6\,M_\odot$.  In contrast to Eq.~\ref{eq:salp},
\be
\alpha\sim -1.2 \quad (M>0.6 M_\odot);\qquad \alpha\sim 0.4\quad
(M<0.6\,M_\odot)\label{eq:realmf}
\ee
Even after correcting for binaries (to which HST is almost completely 
insensitive) the slope at the low-mass end is only $\alpha\sim 0.1$.
There are perhaps hints of a rise in the MF at the very last bin, but the
statistics are too poor to resolve this issue.

\section{Bulge Luminosity Function and Mass Function}

	Light, Baum, \& Holtzman~\cite{lbh} have used the WFC2 to measure
the LF of the galactic bulge in Baade's Window to an apparent magnitude 
$V\sim 26$.
This is not as deep as the images used to measure the disk LF primarily
because the bulge fields are limited by crowding.  Moreover, since the bulge 
is 8 kpc away, while the disk stars can be seen as close as 0.5 kpc, 
(corresponding to an additional factor of 250 in apparent brightness), the
bulge LF measurement is cutoff about 10 magnitudes (factor 10,000 in 
luminosity) brighter than the disk LF.  Even so, this is a factor $\sim 100$
improvement on pre-HST efforts.  The results are noteworthy: to the limit
to which it can be measured, $M_V\sim 10$, the bulge LF coincides with the
disk LF.  Since the heavy-element abundance of bulge stars is similar to those
in the solar neighborhood, the mass-luminosity relation should be similar.
Hence the MFs of the two populations should also be similar.  This suggests
that perhaps the MFs are also the same at the low mass end.  If so, this
leads to some rather dramatic conclusions.

	The dynamically-measured mass of the bulge is 
$\sim 2\times 10^{10}\,M_\odot$.  Han finds that the stars observed by
Light et al.\ account for half of this mass, but can account for no more than
1/10 of the observed microlensing events.~\cite{han}  
If the bulge LF is extended using the disk LF and similarly converted into an 
MF, this would account for 70\% of the bulge mass, but less than 1/2 the
microlensing events and essentially none of the short events.  Only when
Han adds in the remaining 30\% of the mass in brown dwarfs 
($M\sim 0.08\,M_\odot$) can he account for these short events.  In brief,
star count work on the luminous populations seems to suggest that much of
the mass in these components is composed of brown dwarfs or other dark objects
of similar mass.

\section{Hubble Deep Field Search For Halo Stars}

	The Hubble Deep Field (HDF) with a total of 10 days of integration
provides a unique opportunity to probe for extreme halo objects. 
Flynn, Gould, \& Bahcall~\cite{fgb} found that stars could be separated from
galaxies to a limiting magnitude $I=26.3$, about 10 times fainter than
typical WFC2 fields used to measure the disk LF.  Most known populations of
stars in the Galaxy will not generate counts near this faint limit simply
because to do so they would have to be so far away that they would be outside
the Galaxy!  Since the faintest magnitudes reached by HDF are essentially free
of known populations, it can be used to search for objects that are so 
intrinsically faint that they would have escaped notice in earlier studies.
The only ``expected'' candidate of this type are the white dwarfs, for which
HDF give us the first meaningful limits:
\be 
f < 0.31 \times 10^{0.72[(V-I)-1.8]},\label{eq:wdlimit}
\ee
where $f$ is the halo fraction of $0.5\,M_\odot$ white dwarfs and $(V-I)$ is
their color.  Thus, HDF tells us white dwarfs in the expected color range
make up no more than 1/2 to 1/3 of the halo.  More generally, HDF constrains
all classes of objects with absolute magnitude $M_I$, mass $M$, and halo 
fraction $f$ by,
\be
M_I > 17.2 + {5\over 3}\log\biggl(f{0.08\,M_\odot\over M}\biggr)\qquad
(V-I>1.8),\label{eq:alllimit}
\ee
where I have scaled the mass to the maximum of the brown dwarf regime.  This
limit is 10 times fainter than the faintest star ever observed and 100 times
fainter than the faintest halo star ever observed.  In brief, a significant
population (but not the whole halo) of white dwarfs is still permitted, but
ordinary halo stars simply do not contribute to the mass of the Galaxy.

\section{HDF Limits on Intergalactic Stars}

	Intergalactic stars are not often regarded as candidates for dark 
matter, but many cosmological scenarios produce stars at very early times
and these must be distributed approximately as the dark matter.  Thus, it
is of interest to determine their density.  HDF can be used to search
for K giant stars over a volume of about 70 cubic kiloparsecs outside the
Galaxy (but inside the Local Group).  The density is at least a factor
3000 times lower than the local density of giant stars and so more than
300,000 times below the local dark matter density (assuming a locally 
measured MF).  Of course, the Local Group dark matter density is
about 10,000 times lower than the nearby density, so intergalactic
stars make up less than 1/30 of the dark matter in the Local Group.

\section*{Acknowledgments}
This work was supported in part by NSF grant AST 9420746.

\end{document}